\documentclass[12pt]{article}
\usepackage{axodraw,mathrsfs}
\title{\bf Comment on physical-scalar mediated contribution to fermion self energy}
\author{{\bf Biswajit Adhikary\footnote{biswajit.adhikari@saha.ac.in}}\\
   Saha Institute of Nuclear Physics,\\ 1/AF Bidhan  
        Nagar, Kolkata 700064, India }
\date{}

\begin{document}
\maketitle
\begin{abstract}
  We calculate the physical scalar contribution to the fermionic self
  energy matrix at one loop. We make a comment about the
  difference of our results from those in the existing literature.
\end{abstract}
\begin{figure}[htbp]
 \begin{center}\begin{picture}(300,60)(0,45)
\ArrowLine(50,50)(110,50)
\ArrowLine(110,50)(190,50)
\ArrowLine(190,50)(250,50)
\Text(80,55)[b]{$f(p)$}
\Text(220,55)[b]{$f(p)$}
\Text(150,95)[b]{$h_i(k)$}
\DashCArc(150,50)(40,0,180){2}\Vertex(110,50){2}\Vertex(190,50){2}
\end{picture}\end{center}
\caption[]{\label{fen}Self energy diagram of fermion with physical scalar.}\end{figure}
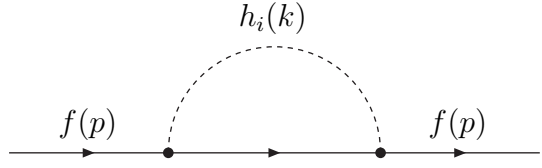

We want to calculate physical scalar contribution to the fermionic
self energy matrix in a generalized gauge theory that was originally
done by Weinberg \cite{wein}. In general there should be a symmetry
breaking scalar potential. Non zero vacuum expectation values of the
scalar fields break the gauge symmetry to lower group and generate
mass terms for gauge fields and fermionic fields. The scalar potential
about the vacuum generate mass term for scalar fields also. The scalar
mass matrix have nonzero eigenvalues along with the zero eigenvalues.
Scalar fields with zero mass are known as Goldstone scalars which can
be absorbed into the gauge fields under proper gauge
transformation. Rest of the scalars are physical.
We have chosen the basis of the scalar fields where they have definite
mass.  We only concentrate on the contribution of those physical
scalars to fermionic self energy matrix at the 1-loop level. The
relevant part of the Lagrangian for our calculations is
\begin{eqnarray}
\mathscr{L}=-\bar{f}_am_{ab}f_b-\sum_i\bar{f}_a(\Gamma_i)_{ab}f_bh_i
\label{lag}
\end{eqnarray}
where $f_a$'s, $h_i$'s, $m$ and $\Gamma_i$'s are respectively the
fermionic fields, physical scalar fields with definite mass $M_i$'s,
zeroth order fermionic mass matrix and Yukawa coupling matrices of the
fermions with the scalars $h_i$'s. The Feynman diagram
relevant for this calculation is in Figure.\ \ref{fen}.
  We choose a basis of the fermionic fields where the fermionic mass
matrix is free from $\gamma_5$. So, it will be hermitian. The
fermionic fields in this basis will be $\hat{f}=Sf$ where $S$ is a
unitary transformation matrix. The fermionic mass matrix, Yukawa
coupling matrix and self energy matrix in this basis will be \cite{gof}
\begin{eqnarray}
\label{mfsb}
\hat{m}=\gamma_0S\gamma_0mS^\dagger,\quad
\hat{\Gamma}_i=\gamma_0S\gamma_0\Gamma_iS^\dagger\quad\mbox{and}\quad\hat{\Sigma}=\gamma_0S\gamma_0\Sigma S^\dagger
\end{eqnarray}
where $\Sigma$ is fermionic self energy matrix in the original basis.
So, the physical scalar contributions to the self energy matrix in
this basis will be
\begin{eqnarray}
i\hat{\Sigma}^{\rm scal}&=&\sum_i\int
\frac{d^4k}{(2\pi)^4}(-i\hat{\Gamma}_i)\times\frac{i}{k^2-M_i^2}\times\frac{i}{\rlap/p-\rlap/k-\hat{m}}\times(-i\hat{\Gamma}_i)
\label{self2}
\end{eqnarray}
The above Eq.\ (\ref{self2}) can be written down as
\begin{eqnarray}
i\hat{\Sigma}^{\rm scal}&=&\sum_i\int
\frac{d^4k}{(2\pi)^4}\hat{\Gamma}_i\times\frac{1}{k^2-M_i^2}\times\frac{\rlap/p-\rlap/ k+\hat{m}}{(p-k)^2-\hat{m}^2}\times\hat{\Gamma}_i.
\label{self3}
\end{eqnarray}
There are basically two fundamental integrals. Their forms in $d$
dimension for
dimensional regularization scheme are
\begin{eqnarray}
&&I=\int \frac{d^dk}{(2\pi)^d}\frac{1}{k^2-M_i^2}\times\frac{1}{(p-k)^2-\hat{m}^2}\nonumber\\
&&I^{\mu}=\int \frac{d^dk}{(2\pi)^d}\frac{1}{k^2-M_i^2}\times\frac{k^\mu}{(p-k)^2-\hat{m}^2}.
\label{int}
\end{eqnarray}
Results of the integrals under Feynman parameterization are
\begin{eqnarray}
&&I=\frac{i}{(4\pi)^{d/2}}\Gamma(\frac{4-d}{2})\int^1_0dx\mathscr{D}^{-(4-d)/2}\nonumber\\
&&I^{\mu}=p^{\mu}\frac{i}{(4\pi)^{d/2}}\Gamma(\frac{4-d}{2})\int^1_0xdx\mathscr{D}^{-(4-d)/2}
\label{intres}
\end{eqnarray}
 where
\begin{eqnarray}
\mathscr{D}=\hat{m}^2x+M_i^2(1-x)-p^2x(1-x).
\label{dform}
\end{eqnarray}
 Using the integral results of Eq.\ (\ref{intres}) in the Eq.\ 
(\ref{self3}) we have
\begin{eqnarray}
i\hat{\Sigma}^{\rm scal}&=&\sum_i(\mu^{(4-d)/2}\hat{\Gamma}_i)\times\{(\rlap/p+\hat{m})I-\gamma_{\mu}I^{\mu}\}\times(\mu^{(4-d)/2}\hat{\Gamma}_i)\nonumber\\
&=&\sum_i(\mu^2)^{(4-d)/2}\hat{\Gamma}_i\frac{i}{(4\pi)^{d/2}}\Gamma(\frac{4-d}{2})\int^1_0dx\{\rlap/p(1-x)+\hat{m}\}\mathscr{D}^{-(4-d)/2}\hat{\Gamma}_i.
\label{self4}
\end{eqnarray}
where $\mu$ is an arbitrary mass scale which has been introduced to keep
$\Gamma_i$ dimensionless in $d$ dimension. Now using the expansion of
type $A^{\epsilon/2}=1+\frac{\epsilon}{2}\ln(A)+O(\epsilon^2)$ and
$\Gamma(\frac{\epsilon}{2})=\frac{2}{\epsilon}-\gamma+O(\epsilon)$
where $\epsilon=4-d$, we obtain
\begin{eqnarray}
i\hat{\Sigma}^{\rm scal}&=&\sum_i\hat{\Gamma}_i\frac{i}{(4\pi)^2}\Big[W\int^1_0dx\{\rlap/p(1-x)+\hat{m}\}\nonumber\\&&-\int^1_0dx\{\rlap/p(1-x)+\hat{m}\}\ln\mathscr{D}+O(\epsilon)\Big]\hat{\Gamma}_i.
\label{self5}
\end{eqnarray}
where 
\begin{eqnarray}
W=\ln(4\pi)+\frac{2}{\epsilon}-\gamma+\ln({\mu}^{2}).
\label{wexp}
\end{eqnarray}

Due to the scalar pseudoscalar bi-linear combination of Yukawa term, $\hat{\Gamma}_i\rlap/p=\rlap/p\gamma_0\hat{\Gamma}_i\gamma_0$.  The self
energy matrix can be written as
\begin{eqnarray}
\Sigma=(\rlap/p -\hat{m})F(p^2)+G(p^2)(\rlap/p -\hat{m})+\Sigma_{\rm eff}(p^{2})
\label{totself}
\end{eqnarray}
Upto  first order term in $F$, $G$ and $\Sigma_{\rm eff}$ fermionic propagator can be written
down as \cite{wein}
\begin{eqnarray}
S_F(p)=\frac{1}{1+G}\times\frac{1}{\rlap/p -\hat{m}+\Sigma_{\rm eff}}\times\frac{1}{1+F}.
\label{prop}
\end{eqnarray}
It shows that the pole of the propagator does not depend on $F$ and
$G$. So, we can easily substitute $\rlap/p$ by $\hat{m}$ whenever
$\rlap/p$ will appear at the extreme right or extreme left in the
expression for $\Sigma$. Following
the above discussion and using $p^2=\hat{m}^2$ inside $\mathscr{D}$
under the consideration of first order correction of mass we
obtain
\begin{eqnarray}
\hat{\Sigma}^{\rm scal}&=&\sum_i\frac{1}{(4\pi)^2}\Big[W\int^1_0dx\{\hat{m}\gamma_0\hat{\Gamma}_i\gamma_0(1-x)+\hat{\Gamma}_i\hat{m}\}\nonumber\\&&-\int^1_0dx\{\hat{m}\gamma_0\hat{\Gamma}_i\gamma_0(1-x)+\hat{\Gamma}_i\hat{m}\}\ln\{\hat{m}^2x^2+M_i^2(1-x)\}+O(\epsilon)\Big]\hat{\Gamma}_i.
\label{self6}
\end{eqnarray}
Coefficient of $W$ is equivalent to the coefficient of
$\ln(\Lambda^2)$ in \cite{wein} where cutoff regularization was used.
Compared to the expression of \cite{wein} this result has different
signs in the terms containing the combination
$\gamma_0\hat{\Gamma}_i\gamma_0$, both in finite as well as diverging
parts. Later various people \cite{gof,Branco} used the results of
Weinberg \cite{wein}. Turning back to the original basis with the
relations in Eq.\ (\ref{mfsb}), using the hermiticity of $\hat m$ and
the decompositions like
$B=B_R(1+\gamma_5)/2+B_R^\dagger(1-\gamma_5)/2$ for both $m$ and
$\Gamma_i$ due to the hermiticity of the Lagrangian we obtain
\begin{eqnarray}
\Sigma^{\rm scal}&=&\sum_i\frac{1}{(4\pi)^2}\Big[W\int^1_0dx\{m\Gamma_i^\dagger(1-x)+\Gamma_im^\dagger\}\nonumber\\&&-\int^1_0dx\{m\Gamma_i^\dagger(1-x)+\Gamma_im^\dagger
\}\ln\{mm^\dagger x^2+M_i^2(1-x)\}+O(\epsilon)\Big]\Gamma_i.
\end{eqnarray}
which is similar to the results of \cite{gof} except the sign
correction here. \\
{\bf Note Added}: After doing the above calculations we came to know
that in another paper \cite{bordes1} the authors used the results of
\cite{wein}. Later \cite{bordes} it was commented that there was a
sign error in \cite{bordes1}.
\paragraph*{Acknowledgments:}
The author thanks Palash B. Pal for reading the manuscript and
suggesting improvements.

 \end{document}